\documentclass[article,
 amsmath,amssymb,
 aps,
]{revtex4-2}

\usepackage{graphicx}
\usepackage{dcolumn}
\usepackage{bm}
\usepackage{xcolor}

\begin{document}


\title{The consequences of non-differentiable angular dispersion in optics: Tilted pulse fronts versus space-time wave packets}

\author{Layton A. Hall$^{1}$}
\author{Ayman F. Abouraddy$^{1,*}$}
\affiliation{$^{1}$CREOL, The College of Optics \& Photonics, University of Central~Florida, Orlando, FL 32816, USA}
\affiliation{$^*$Corresponding author: raddy@creol.ucf.edu}

\begin{abstract}
Conventional diffractive and dispersive devices introduce angular dispersion (AD) into pulsed optical fields thus producing so-called `tilted pulse fronts'. Naturally, it is always assumed that the functional form of the wavelength-dependent propagation angle associated with AD is differentiable with respect to wavelength. Recent developments in the study of space-time wave packets -- pulsed beams in which the spatial and temporal degrees of freedom are inextricably intertwined -- have pointed to the existence of \textit{non-differentiable} AD: field configurations in which the propagation angle does not possess a derivative at some wavelength. Here we investigate the consequences of introducing non-differentiable AD into a pulsed field and show that it is the crucial ingredient required to realize group velocities that deviate from $c$ (the speed of light in vacuum) along the propagation axis in free space. In contrast, the on-axis phase and group velocities are always equal in conventional scenarios. Furthermore, we show that non-differentiable AD is needed for realizing anomalous or normal group-velocity dispersion along the propagation axis, while simultaneously suppressing all higher-order dispersion terms. These and several other consequences of non-differentiable AD are verified experimentally using a pulsed-beam shaper capable of introducing AD with arbitrary spectral profile. Rather than being an exotic phenomenon, non-differentiable AD is an accessible, robust, and versatile resource for sculpting pulsed optical fields.
\end{abstract}


\maketitle

\section{Introduction}

Whenever a plane-wave pulse, in which all the frequencies travel in the same direction, traverses a diffractive or dispersive device, such as a diffraction grating or a prism, angular dispersion (AD) is introduced into the field structure: each frequency $\omega$ now travels at a different angle $\varphi(\omega)$ with respect to the propagation axis. After having its genesis in Newton's experiments on the spectral analysis of light with prisms \cite{Sabra81Book}, AD has become a mainstay in laser physics \cite{Fulop10Review,Torres10AOP} with applications in traveling-wave optical amplification \cite{Bor83APB,Klebniczki88APB,Hebling89OL,Hebling91JOSAB}, dispersion compensation \cite{Martinez84JOSAA,Fork84OL,Gordon84OL,Szatmari96OL}, pulse compression \cite{Bor85OC,Lemoff93OL,Kane97JOSAB,Kane97JOSAB2}, broadening the phase-matching bandwidth in nonlinear optics \cite{Martinez89IEEE,Szabo90APB,Szabo94APB,Richman98OL,Richman99AO}, and in the generation of THz pulses \cite{Hebling02OE,Nugraha19OL,Wang20LPR}.

The field structure resulting from AD is usually called a `tilted-pulse front' (TPF) because the pulse front is tilted with respect to the phase front, leading to a characteristic spatio-temporal profile that is tilted with respect to the propagation axis \cite{Bor93OE,Hebling96OQE}. In the conventional theory of AD \cite{Fulop10Review,Torres10AOP,Porras03PRE2}, the propagation angle $\varphi(\omega)$ is expanded in a perturbative series at a carrier frequency $\omega_{\mathrm{o}}$, which of course presumes that $\varphi(\omega)$ is differentiable at $\omega\!=\!\omega_{\mathrm{o}}$. This appears to be a natural premise that need not be questioned, especially in light of the form of AD produced by conventional optical components \cite{Fork84OL,Gordon84OL,Porras03PRE2,Szatmari96OL,Fulop10Review,Torres10AOP}. However, recent work on `space-time' (ST) wave packets \cite{Kondakci16OE,Parker16OE,Kondakci17NP,Wong17ACSP2,Porras17OL,Efremidis17OL,Yessenov19OPN}, pulsed optical beams in which the spatial and temporal degrees of freedom are inextricably intertwined \cite{Donnelly93ProcRSLA,Saari04PRE,Longhi04OE,Wong17ACSP1,Kondakci18PRL,Kondakci19OL,Yessenov19PRA,Yessenov19OE,Wong20AS,Wong21OE}, has led us to re-examine this time-tested assumption. The initial grounds for this re-examination was the observation that ST wave packets are themselves endowed with AD and thus have a similar overall tilted pulse-front structure \cite{Wong17ACSP2,Kondakci19ACSP}, and yet some of their properties nevertheless depart qualitatively from those well-established for TPFs. For example, a defining characteristics of a TPF is that the tilt angle of its pulse front is related to the underlying AD by a universal relationship that is independent of the pulse bandwidth \cite{Bor93OE,Hebling96OQE}. In contrast, we recently demonstrated experimentally that the tilt angle of a ST wave-packet pulse front depends inversely on the square-root of the bandwidth \cite{Hall21OL}. We have identified this feature along with other unique properties of ST wave packets as being a consequence of \textit{non-differentiable} AD undergirding the field. That is, the assumption underlying conventional AD fails in the case of ST wave packets, and the propagation angle $\varphi(\omega)$ is \textit{not} differentiable at the frequency $\omega\!=\!\omega_{\mathrm{o}}$ \cite{Hall21OL,Yessenov21ACSP,Hall21OL3NormalGVD}. We emphasize that non-differentiable AD is \textit{not} an exotic nor a pathological condition, but is instead one that can be readily produced and tuned experimentally, albeit not with conventional optics devices commonly used to introduce AD, but with the pulsed-beam shaper we introduced for synthesizing ST wave packets \cite{Kondakci17NP,Kondakci18PRL,Kondakci19NC,Yessenov19OE,Bhaduri19Optica,Bhaduri20NP}. 

Here we examine in more detail the consequences of introducing non-differentiable AD into a pulsed optical field. Starting from the basic theoretical formulation of AD, we show that non-differentiable AD leads to multiple consequences that contrast with traditional expectations. First, in presence of non-differentiable AD the group velocity $\widetilde{v}$ can differ in free space from the phase velocity $v_{\mathrm{ph}}$ along the propagation axis; in contrast, we always have $v_{\mathrm{ph}}\!=\!\widetilde{v}\!=\!c$ ($c$ is the speed of light in vacuum) along the propagation axis in TPFs \cite{Porras03PRE2}. Second, as mentioned above, the pulse-front tilt angle is bandwidth-dependent in the vicinity of $\omega_{\mathrm{o}}$ in presence of non-differentiable AD, but is independent of bandwidth in presence of differentiable AD \cite{Hall21OL}. Third, non-differentiable AD allows for wide tunability of $\widetilde{v}$ along the propagation axis \cite{Kondakci19NC,Yessenov19PRA,Yessenov19OL}. Fourth, TPFs endowed with differentiable AD are always accompanied by GVD \cite{Martinez84JOSAA,Porras03PRE2,Torres10AOP}, whereas ST wave packets endowed with non-differentiable AD can be GVD-free \cite{Kondakci17NP,Kondakci18PRL}. Although gratings can produce significant deviations in $\widetilde{v}$ from $c$, these occur only at large diffraction angles rather than on-axis as in the case of non-differentiable AD and are accompanied by strong GVD \cite{Porras03PRE2}. Fifth, non-differentiable AD can yield either normal or anomalous GVD in free space along the propagation axis \cite{Yessenov21ACSP,Hall21OL3NormalGVD}, in contradistinction to a well-known result in laser physics that indicates that (differentiable) AD yields \textit{only} anomalous GVD \cite{Martinez84JOSAA}. Sixth, non-differentiable AD facilitates introducing arbitrary dispersion profiles along the propagation direction, including profiles truncated at a particular order while suppressing all higher-order dispersion terms \cite{Yessenov21ACSP}.

Using a pulsed-beam shaper that can readily introduce differentiable or non-differentiable AD into a pulsed plane-wave, we verify \textit{all} of these predicted phenomena experimentally. From our analysis and measurements we make the following distinction between TPFs and ST wave packets: TPFs are pulsed optical fields endowed with differentiable AD and the conventional features associated with AD are therefore expected to be observed; whereas ST wave packets are endowed instead with non-differentiable AD, and many of the traditional expectations are thus overturned. Non-differentiable AD is therefore a robust and versatile resource that can have dramatically alter the propagation characteristics of the field, and may lead to new applications in dispersion compensation, nonlinear optics, and optical signal processing.

\section{Theoretical Formulation}

\subsection{Perturbative treatment of angular dispersion}

We start from a Taylor expansion of the propagation angle $\varphi(\omega)$ with respect to the $z$-axis:
\begin{equation}\label{Eq:TaylorExpansionOfTheAngle}
\varphi(\omega)=\varphi(\omega_{\mathrm{o}}+\Omega)=\varphi_{\mathrm{o}}+\varphi_{\mathrm{o}}^{(1)}\Omega+\tfrac{1}{2}\varphi_{\mathrm{o}}^{(2)}\Omega^{2}+\cdots,
\end{equation}
where $\Omega\!=\!\omega-\omega_{\mathrm{o}}$, $\omega_{\mathrm{o}}$ is the carrier frequency, $\varphi_{\mathrm{o}}\!=\!\varphi(\omega_{\mathrm{o}})$, $\varphi_{\mathrm{o}}^{(n)}\!=\!\tfrac{d^{n}\varphi}{d\omega^{n}}\big|_{\omega=\omega_{\mathrm{o}}}$, and the subscript `o' refers to quantities evaluated at $\omega_{\mathrm{o}}$. The set of coefficients $\{\varphi_{\mathrm{o}}^{(n)}\}\!=\!\{\varphi_{\mathrm{o}},\varphi_{\mathrm{o}}^{(1)},\varphi_{\mathrm{o}}^{(2)},\cdots\}$ are referred to as the AD coefficients. Rotating the coordinate system to align the frequency $\omega_{\mathrm{o}}$ with the $z$-axis ($\varphi_{\mathrm{o}}\!=\!0$) does not impact the AD coefficients $\varphi_{\mathrm{o}}^{(n)}$ for $n\!\geq\!1$. It suffices to consider one transverse coordinate $x$, and the transverse $k_{x}(\omega)\!=\!k\sin\{\varphi(\omega)\}$ and axial $k_{z}(\omega)\!=\!k\cos\{\varphi(\omega)\}$ wave numbers can each be expanded in a Taylor series ($k\!=\!\omega/c$ and $k_{\mathrm{o}}\!=\!\omega_{\mathrm{o}}/c$):
\begin{equation}
k_{x}(\omega)=k_{x}^{(0)}+k_{x}^{(1)}\Omega+\tfrac{1}{2}k_{x}^{(2)}\Omega^{2}+\cdots,\;\;\;k_{z}(\omega)=k_{z}^{(0)}+k_{z}^{(1)}\Omega+\tfrac{1}{2}k_{z}^{(2)}\Omega^{2}+\cdots,
\end{equation}
where we refer to $\{k_{x}^{(n)}\}$ and $\{k_{z}^{(n)}\}$ as the transverse and axial dispersion coefficients, respectively. In free space, the dispersion coefficients are determined solely by the AD coefficients $\{\varphi_{\mathrm{o}}^{(n)}\}$ \cite{Porras03PRE2}:
\begin{eqnarray}
k_{x}^{(0)}\!\!\!&=&\!\!k_{\mathrm{o}}\sin{\varphi_{\mathrm{o}}},\\k_{z}^{(0)}\!\!\!&=&\!\!k_{\mathrm{o}}\cos{\varphi_{\mathrm{o}}},\\
ck_{x}^{(1)}\!\!\!&=&\omega_{\mathrm{o}}\varphi_{\mathrm{o}}^{(1)}\cos{\varphi_{\mathrm{o}}}+\sin{\varphi_{\mathrm{o}}},\\ ck_{z}^{(1)}\!\!\!&=&\!\!\cos{\varphi_{\mathrm{o}}}-\omega_{\mathrm{o}}\varphi_{\mathrm{o}}^{(1)}\sin{\varphi_{\mathrm{o}}},\label{Eq:AxialFirstOrderWaveNumber}\\
c\omega_{\mathrm{o}}k_{x}^{(2)}\!\!\!&=&\!\!(\omega_{\mathrm{o}}^{2}\varphi_{\mathrm{o}}^{(2)}+2\omega_{\mathrm{o}}\varphi_{\mathrm{o}}^{(1)})\cos{\varphi_{\mathrm{o}}}-(\omega_{\mathrm{o}}\varphi_{\mathrm{o}}^{(1)})^{2}\sin{\varphi_{\mathrm{o}}},\label{Eq:TransverseGVDCoefficient}\\
c\omega_{\mathrm{o}}k_{z}^{(2)}\!\!\!&=&\!\!-(\omega_{\mathrm{o}}\varphi_{\mathrm{o}}^{(1)})^{2}\cos{\varphi_{\mathrm{o}}}-(\omega_{\mathrm{o}}^{2}\varphi_{\mathrm{o}}^{(2)}+2\omega_{\mathrm{o}}\varphi_{\mathrm{o}}^{(1)})\sin{\varphi_{\mathrm{o}}}\label{Eq:AxialGVDCoefficient}.
\end{eqnarray}

\begin{figure}[t!]
\centering
\includegraphics[width=11cm]{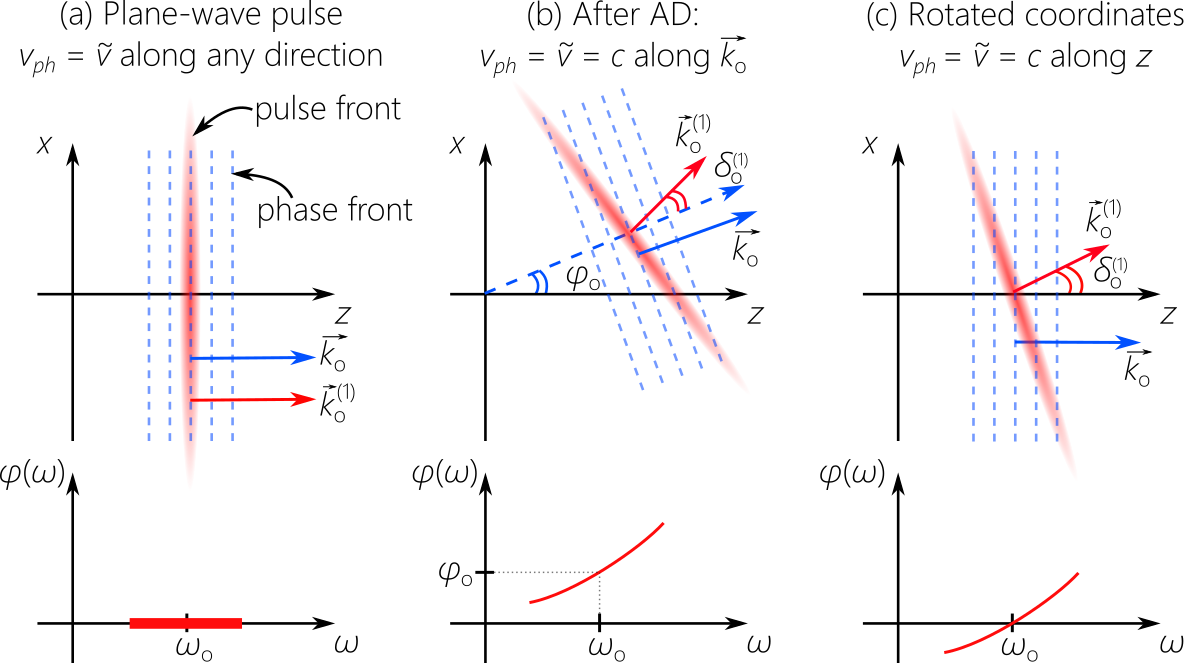}
\caption{\textit{Differentiable} AD for TPFs. (a) In a plane-wave pulse, $\vec{k}_{\mathrm{o}}$ and $\vec{k}_{\mathrm{o}}^{(1)}$ co-align, the phase front and pulse front coincide, and $v_{\mathrm{ph}}\!=\!\widetilde{v}$ along any direction. (b) After introducing AD, each frequency $\omega$ travels at a different angle $\varphi(\omega)$, $\vec{k}_{\mathrm{o}}$ makes an angle $\varphi_{\mathrm{o}}$ with the $z$-axis, $\vec{k}_{\mathrm{o}}^{(1)}$ an angle $\delta_{\mathrm{o}}^{(1)}$ with $\vec{k}_{\mathrm{o}}$, the pulse front is tilted with respect to the phase front, $v_{\mathrm{ph}}\!=\!\widetilde{v}\!=\!c$ along $\vec{k}_{\mathrm{o}}$, but $v_{\mathrm{ph}}\!\neq\!\widetilde{v}$ along any other direction. (c) Rotating the coordinate system in (b) to align $\vec{k}_{\mathrm{o}}$ with the $z$-axis. Under each panel we plot $\varphi(\omega)$.}
\label{Fig:DifferentiableAngularDispersion}
\end{figure}

\subsection{Phase velocity}

All the frequencies in an AD-free plane-wave pulse travel in the same direction [Fig.~\ref{Fig:DifferentiableAngularDispersion}(a)], the phase velocity is luminal $v_{\mathrm{ph}}\!=\!c$ along the propagation direction, and the phase front (the plane of constant phase) coincides with the pulse front (the plane of constant amplitude). After introducing AD via a grating or a prism, for example, each frequency $\omega$ travels at a different angle $\varphi(\omega)$. The phase front is orthogonal to the vector $\vec{k}_{\mathrm{o}}\!=\!(k_{x}^{(0)},k_{z}^{(0)})\!=\!k_{\mathrm{o}}(\sin{\varphi_{\mathrm{o}}},\cos{\varphi_{\mathrm{o}}})$. Although we still have $v_{\mathrm{ph}}\!=\!c$ along $\vec{k}_{\mathrm{o}}$, along the $z$-axis we have $v_{\mathrm{ph}}\!=\!\tfrac{\omega_{\mathrm{o}}}{k_{z}^{(0)}}\!=\!\tfrac{c}{\cos{\varphi_{\mathrm{o}}}}$ [Fig.~\ref{Fig:DifferentiableAngularDispersion}(a)]. The deviation of $v_{\mathrm{ph}}$ along $z$ from $c$ in this case is a well-known geometric effect due to the tilt of the phase front with respect to the observation axis along $z$ \cite{Chiao02OPN}. Note that $v_{\mathrm{ph}}$ depends only on $\varphi_{\mathrm{o}}$, and is independent of all higher-order AD terms. Of course, by aligning $\vec{k}_{\mathrm{o}}$ with the $z$-axis, we once again have $v_{\mathrm{ph}}\!=\!c$ along $z$ [Fig.~\ref{Fig:DifferentiableAngularDispersion}(c)].

\subsection{Group velocity}

In presence of AD, the \textit{pulse front} is orthogonal to the vector $\vec{k}_{\mathrm{o}}^{(1)}\!=\!(k_{x}^{(1)},k_{z}^{(1)})$, which makes an angle $\delta_{\mathrm{o}}^{(1)}$ with $\vec{k}_{\mathrm{o}}$. This tilt angle is given by the universal relationship:
\begin{equation}
\tan{\delta_{\mathrm{o}}^{(1)}}=\omega_{\mathrm{o}}\varphi_{\mathrm{o}}^{(1)},
\end{equation}
which is independent of the pulse width and shape, and is independent of the device inculcating AD into the field \cite{Hebling96OQE}; see Fig.~\ref{Fig:DifferentiableAngularDispersion}(b). The field thus takes on the form of a TPF \cite{Fulop10Review,Torres10AOP}. The group velocity along the $z$-axis $\widetilde{v}\!=\!\tfrac{1}{k_{z}^{(1)}}$ from Eq.~\ref{Eq:AxialFirstOrderWaveNumber} is: 
\begin{equation}\label{Eq:GroupVelocityGeneral}
\widetilde{v}=\frac{c}{\cos{\varphi_{\mathrm{o}}}-\omega_{\mathrm{o}}\varphi_{\mathrm{o}}^{(1)}\sin{\varphi_{\mathrm{o}}}}=c\frac{\cos{\delta_{\mathrm{o}}^{(1)}}}{\cos{(\delta_{\mathrm{o}}^{(1)}+\varphi_{\mathrm{o}})}}.
\end{equation}
In \textit{absence} of AD $\varphi_{\mathrm{o}}^{(1)}\!=\!0$, $\widetilde{v}\!=\!v_{\mathrm{ph}}$ along any direction [Fig.~\ref{Fig:DifferentiableAngularDispersion}(a)]. In \textit{presence} of AD $\varphi_{\mathrm{o}}^{(1)}\!\neq\!0$, $\widetilde{v}\!=\!v_{\mathrm{ph}}\!=\!c$ along $\vec{k}_{\mathrm{o}}$, but $\widetilde{v}\!\neq\!v_{\mathrm{ph}}$ along any other direction [Fig.~\ref{Fig:DifferentiableAngularDispersion}(b)]. Aligning $\vec{k}_{\mathrm{o}}$ with the $z$-axis results in $\widetilde{v}\!=\!v_{\mathrm{ph}}\!=\!c$ along $z$ [Fig.~\ref{Fig:DifferentiableAngularDispersion}(c)]. We show in the next Section that these seemingly unavoidable results are overturned in the presence of non-differentiable AD.

\begin{figure}[t!]
\centering
\includegraphics[width=13.2cm]{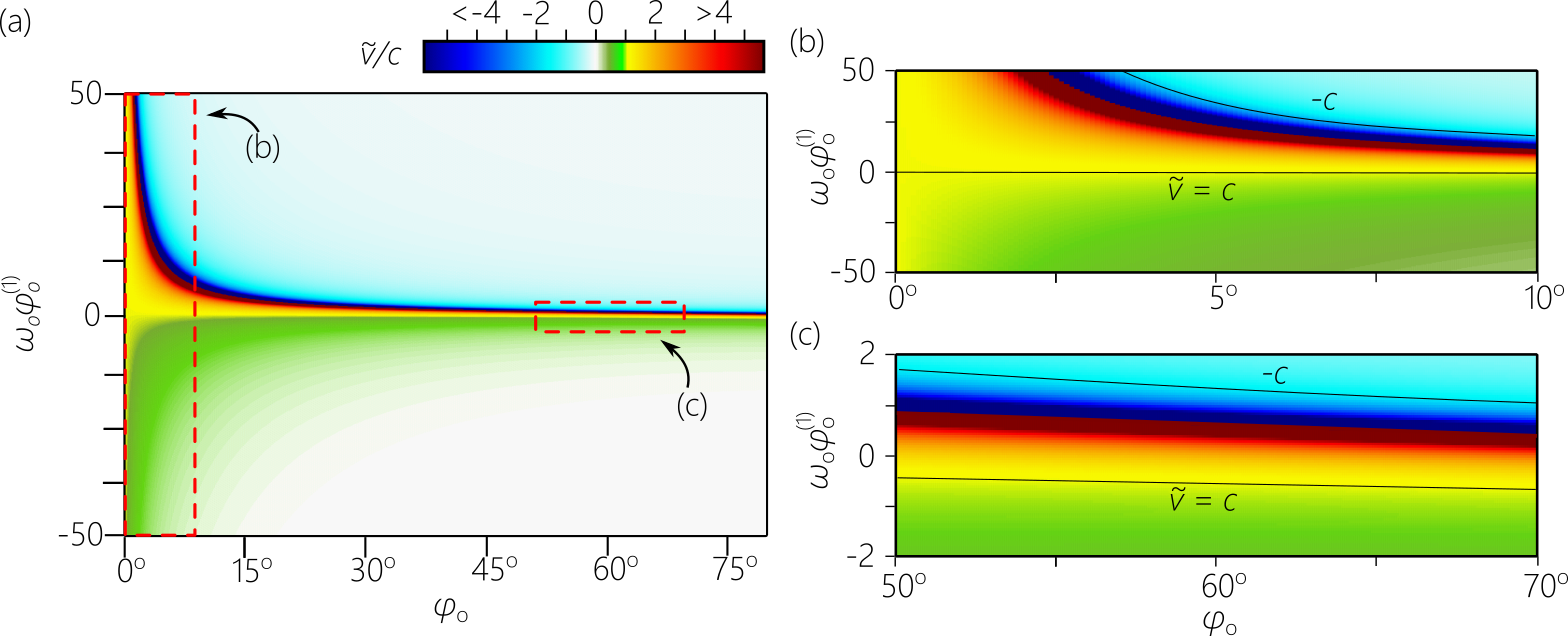}
\caption{(a) Group velocity $\widetilde{v}$ as a function of $\varphi_{\mathrm{o}}$ and $\omega_{\mathrm{o}}\varphi_{\mathrm{o}}^{(1)}$ according to Eq.~\ref{Eq:GroupVelocityGeneral}. (b,c) Plots of $\widetilde{v}$ as a function of $\varphi_{\mathrm{o}}$ and $\omega_{\mathrm{o}}\varphi_{\mathrm{o}}^{(1)}$ corresponding to the two dashed boxes in (a).}
\label{Fig:GroupVelocity}
\end{figure}

We plot in Fig.~\ref{Fig:GroupVelocity}(a) $\widetilde{v}$ from Eq.~\ref{Eq:GroupVelocityGeneral} with $\varphi_{\mathrm{o}}$ and $\omega_{\mathrm{o}}\varphi_{\mathrm{o}}^{(1)}$. When $\varphi_{\mathrm{o}}$ and $\omega_{\mathrm{o}}\varphi_{\mathrm{o}}^{(1)}$ can be tuned independently of each other, the subluminal, luminal, superluminal, and negative-$\widetilde{v}$ regimes can be spanned. Two distinct extremes emerge: for large values of $\omega_{\mathrm{o}}\varphi_{\mathrm{o}}^{(1)}\!\sim\!50$, one can sweep across the full span of values of $\widetilde{v}$ at small propagation angles $\varphi_{\mathrm{o}}$ [Fig.~\ref{Fig:GroupVelocity}(b)], whereas small values of $\omega_{\mathrm{o}}\varphi_{\mathrm{o}}^{(1)}\!\sim\!1$ necessitate large $\varphi_{\mathrm{o}}$ to sweep across comparable values of $\widetilde{v}$ [Fig.~\ref{Fig:GroupVelocity}(c)]. The former scenario is preferable for fields that are required to propagate for significant distances, whereas the latter fields are only useful for interaction with localized structures. Incidentally, the AD produced by gratings follows an opposite trend: large $\omega_{\mathrm{o}}\varphi_{\mathrm{o}}^{(1)}$ is typically produced only at extreme angles $\varphi_{\mathrm{o}}\!\rightarrow\!90^{\circ}$, and small $\omega_{\mathrm{o}}\varphi_{\mathrm{o}}^{(1)}$ is usually produced at small $\varphi_{\mathrm{o}}$. We show below that the pulsed-beam shaper that we recently developed for synthesizing ST wave packets \cite{Kondakci17NP,Kondakci19NC,Bhaduri19Optica,Yessenov19OE,Yessenov19OPN,Bhaduri20NP} is capable of reaching the requisite large values of $\omega_{\mathrm{o}}\varphi_{\mathrm{o}}^{(1)}$ on-axis $\varphi_{\mathrm{o}}\!\approx\!0$.

\subsection{Group-velocity dispersion (GVD)}

\begin{figure}[t!]
\centering
\includegraphics[width=7cm]{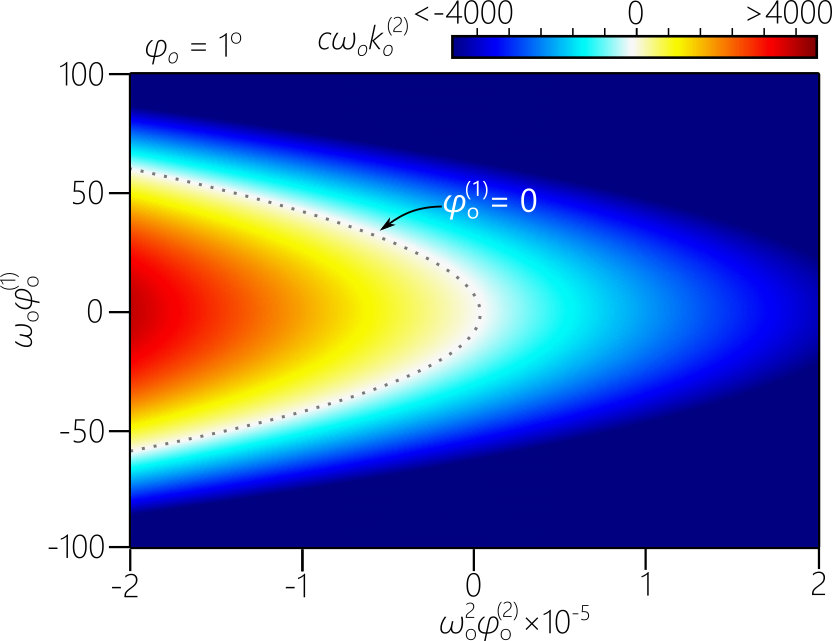}
\caption{Axial GVD coefficient as a function of $\omega_{\mathrm{o}}\varphi_{\mathrm{o}}^{(1)}$ and $\omega_{\mathrm{o}}^{2}\varphi_{\mathrm{o}}^{(2)}$ at $\varphi_{\mathrm{o}}\!=\!1^{\circ}$ (Eq.~\ref{Eq:AxialGVDCoefficient}).}
\label{Fig:GVD}
\end{figure}

The axial and transverse GVD coefficients resulting from AD are given in Eq.~\ref{Eq:TransverseGVDCoefficient} and Eq.~\ref{Eq:AxialGVDCoefficient}, respectively. At $\varphi_{\mathrm{o}}\!=\!0$ ($\vec{k}_{\mathrm{o}}$ is aligned with the $z$-axis): $c\omega_{\mathrm{o}}k_{x}^{(2)}\!=\!\omega_{\mathrm{o}}^{2}\varphi_{\mathrm{o}}^{(2)}+2\omega_{\mathrm{o}}\varphi_{\mathrm{o}}^{(1)}$ and $c\omega_{\mathrm{o}}k_{z}^{(2)}\!=\!-(\omega_{\mathrm{o}}\varphi_{\mathrm{o}}^{(1)})^{2}$. Crucially, the axial GVD in this case is always \textit{anomalous} $k_{z}^{(2)}\!<\!0$ \cite{Martinez84JOSAA}. It must be emphasized that this result applies along only $\vec{k}_{\mathrm{o}}$ where $\widetilde{v}\!=\!v_{\mathrm{ph}}\!=\!c$. Two different strategies can be followed to overturn this dictum, and thus produce anomalous as well as normal GVD, the first of which concerns \textit{off-axis} field configurations $\varphi_{\mathrm{o}}\!\neq\!0$ and the other concerns \textit{on-axis} fields $\varphi_{\mathrm{o}}\!=\!0$.

First, for off-axis fields Eq.~\ref{Eq:AxialGVDCoefficient} indicates that the axial GVD may be either normal or anomalous with judicious control over the three lowest-order AD coefficients $\varphi_{\mathrm{o}}$, $\varphi_{\mathrm{o}}^{(1)}$, and $\varphi_{\mathrm{o}}^{(2)}$. An example was suggested theoretically by Porras \textit{el al}. in \cite{Porras03PRE2} by setting $\varphi_{\mathrm{o}}^{(1)}\!=\!0$, whereupon $c\omega_{\mathrm{o}}k_{z}^{(2)}\!=\!-\omega_{\mathrm{o}}^{2}\varphi_{\mathrm{o}}^{(2)}\sin{\varphi_{\mathrm{o}}}$, so that adjusting the signs of $\varphi_{\mathrm{o}}$ and $\varphi_{\mathrm{o}}^{(2)}$ yields either normal or anomalous GVD. More generally, we plot in Fig.~\ref{Fig:GVD} the GVD coefficient $c\omega_{\mathrm{o}}k_{z}^{(2)}$ as a function of the AD coefficients $\omega_{\mathrm{o}}\varphi_{\mathrm{o}}^{(1)}$ and $\omega_{\mathrm{o}}^{2}\varphi_{\mathrm{o}}^{(2)}$ at $\varphi\!=\!1^{\circ}$. From Fig.~\ref{Fig:GVD} it is clear that either normal or anomalous GVD can be realized at fixed $\varphi_{\mathrm{o}}$ if $\varphi_{\mathrm{o}}^{(1)}$ and $\varphi_{\mathrm{o}}^{(2)}$ can be tuned independently. Unfortunately, conventional devices such as gratings and prisms do not provide such control. Consequently, only anomalous GVD has been produced by AD to date. The theoretical proposal in \cite{Porras03PRE2} to produce normal GVD via AD has only been put to test and verified very recently \cite{Hall21OL3NormalGVD}. It is important to note however that that higher-order dispersion terms inevitably arise and accompany the targeted GVD.

Crucially, the differentiable AD associated with TPFs does \textit{not} provide the possibility for producing normal GVD in on-axis fields. In this case, it seems that the classic result in \cite{Martinez84JOSAA} is unassailable. We show below that this apparently insurmountable barrier is bridged over in the presence of non-differentiable AD associated with ST wave packets, and normal GVD can be exhibited even by on-axis fields. 

\subsection{Summary of the consequences of differentiable AD for TPFs}

We summarize the relevant consequences of conventional differentiable AD for TPFs as follows:
\begin{enumerate}
    \item Along the propagation axis $v_{\mathrm{ph}}\!=\!\widetilde{v}\!=\!c$.
    \item The pulse front is tilted with respect to the pulse front by a bandwidth-independent angle $\delta_{\mathrm{o}}^{(1)}$, where $\tan{\delta_{\mathrm{o}}^{(1)}}\!=\!\omega_{\mathrm{o}}\varphi_{\mathrm{o}}^{(1)}$.
    \item Large deviations of $\widetilde{v}$ from $c$ typically occur at large $\varphi_{\mathrm{o}}$ and $\varphi_{\mathrm{o}}^{(1)}$.
    \item GVD always accompanies AD.
    \item Only anomalous GVD can be produced in free space along the propagation axis.
    \item Higher-order dispersion terms accompany the GVD.
\end{enumerate}
All these results are well-justified theoretically, and have been borne out in experiments over the past 5 decades of research in laser physics. We now proceed to demonstrate theoretically and experimentally that non-differentiable AD helps overturn \textit{all} of these traditional expectations.

\section{Non-differentiable angular dispersion}

We now examine the case where $\varphi(\omega)$ does \textit{not} have a derivative at $\omega_{\mathrm{o}}$, a configuration we refer to as \textit{non-differentiable AD}, and we refer to $\omega_{\mathrm{o}}$ as the non-differentiable frequency.

\subsection{Phase velocity , group velocity and propagation invariance}

Because $v_{\mathrm{ph}}$ depends only on $\varphi_{\mathrm{o}}$, it is independent of the differentiability of $\varphi(\omega)$, in contrast to $\widetilde{v}$ that includes the derivative of $\varphi(\omega)$. Without loss of generality, we align $\vec{k}_{\mathrm{o}}$ with the $z$-axis, so that $\widetilde{v}\!=\!v_{\mathrm{ph}}\!=\!c$ along $z$ [Fig.~\ref{Fig:DifferentiableAngularDispersion}(c)]. To demonstrate that the non-differentiability of $\varphi(\omega)$ at $\omega_{\mathrm{o}}$ can yield $\widetilde{v}\!\neq\!c$, we consider a frequency $\omega$ close to $\omega_{\mathrm{o}}$ associated with the propagation angle $\varphi\!\rightarrow\!0$; $\sin{\varphi}\!\approx\!\varphi$, $\cos{\varphi}\!\approx\!1$, and $\widetilde{v}(\omega)\!\approx\!c/(1-\omega\varphi\tfrac{d\varphi}{d\omega})$. The product $\varphi\tfrac{d\varphi}{d\omega}$ may \textit{not} tend to 0 if $\tfrac{d\varphi}{d\omega}$ diverges at $\omega_{\mathrm{o}}$. Indeed, setting $\varphi\tfrac{d\varphi}{d\omega}\!=\!\tfrac{1}{2\omega_{\mathrm{o}}}\eta^{2}$ to be a frequency-independent constant , where $\eta$ is a dimensionless parameter, and then integrating yields:
\begin{equation}\label{Eq:SquareRootOmega}
\varphi(\omega)=\eta\sqrt{\frac{\Omega}{\omega_{\mathrm{o}}}},
\end{equation}
which is \textit{not} differentiable at $\omega_{\mathrm{o}}$. Substitution into Eq.~\ref{Eq:GroupVelocityGeneral} yields $\widetilde{v}\!=\!c/\widetilde{n}\!\neq\!c$, where $\widetilde{n}\!=\!1-\tfrac{1}{2}\eta^{2}$ is the effective group index \cite{Kondakci19NC,Yessenov19OE,Yessenov20NC}. Crucially, $\widetilde{v}$ is independent of $\omega$, and the dispersion-free wave packet, $k_{z}^{(n)}\!=\!0$ for $n\!\geq\!2$, travels rigidly along $z$ with no temporal broadening. This analysis applies to the non-differentiable frequency $\omega_{\mathrm{o}}$. In its vicinity, $\varphi(\omega)$ is differentiable, but $\omega\tfrac{d\varphi}{d\omega}$ changes rapidly such that $\widetilde{v}\!=\!c/\widetilde{n}$ is maintained. In general, retaining the frequency-independence of $\widetilde{v}$ requires that $k_{z}\!=\!\tfrac{\omega}{c}\cos\{\varphi(\omega)\}\!=\!k_{\mathrm{o}}+\tfrac{\Omega}{\widetilde{v}}$ \cite{FigueroaBook14,Kondakci17NP,Yessenov19PRA}, so that:
\begin{equation}
\sin{\varphi}=\eta\sqrt{\frac{\Omega}{\omega_{\mathrm{o}}}}\;\frac{\omega_{\mathrm{o}}}{\omega}\;\sqrt{1+\frac{1+\widetilde{n}}{2}\frac{\Omega}{\omega_{\mathrm{o}}}},
\end{equation}
which tends to the approximate expression in Eq.~\ref{Eq:SquareRootOmega} when $\Omega\!\ll\!\omega_{\mathrm{o}}$.

In the superluminal regime $\widetilde{v}\!>\!c$, only positive $\Omega\!>\!0$ ($\omega\!>\!\omega_{\mathrm{o}}$) are allowed and $\omega_{\mathrm{o}}$ is the minimum admissible frequency. In the subluminal regime $\widetilde{v}\!<\!c$, only negative $\Omega\!<\!0$ ($\omega\!<\!\omega_{\mathrm{o}}$) are allowed, and $\omega_{\mathrm{o}}$ is the maximum admissible frequency.

\begin{figure}[t!]
\centering
\includegraphics[width=11cm]{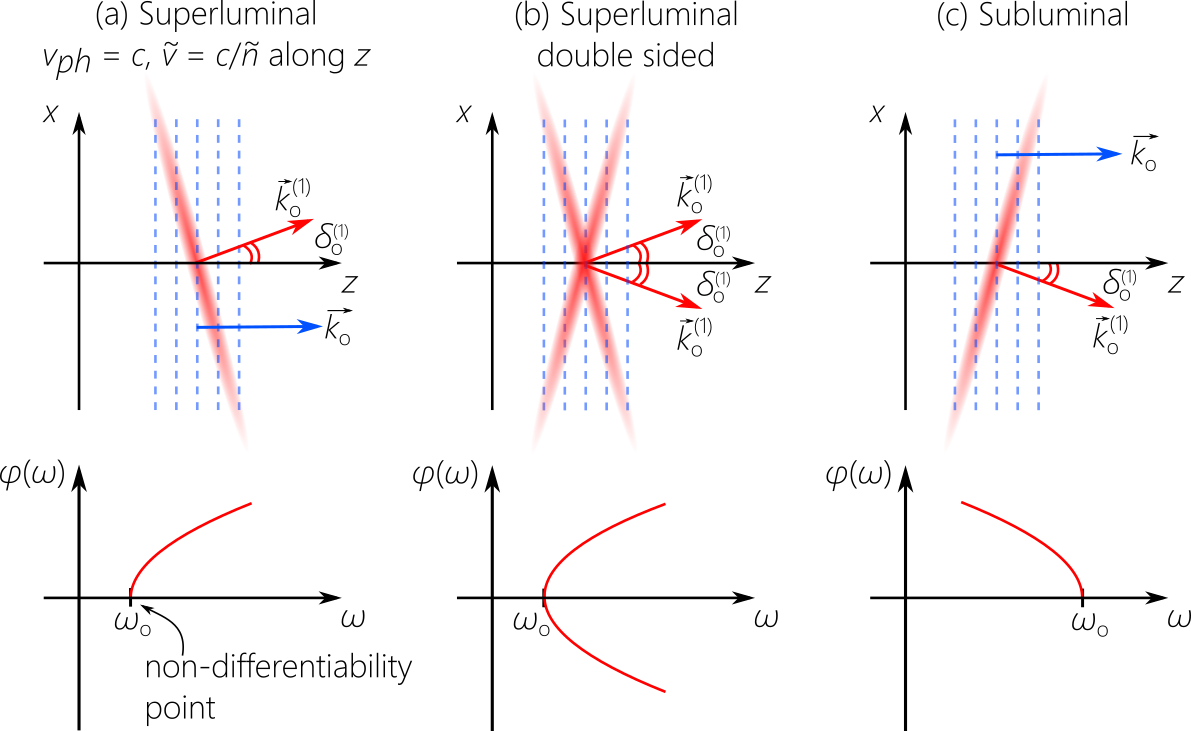}
\caption{\textit{Non-differentiable} AD for ST wave packets. (a) A superluminal ST wave packet incorporating non-differentiable AD at $\omega\!=\!\omega_{\mathrm{o}}$ (the minimum admissible frequency in the spectrum). Similar to a TPF, the phase front is orthogonal to $\vec{k}_{\mathrm{o}}$ that is aligned with the $z$-axis, and the pulse front is orthogonal to $\vec{k}_{\mathrm{o}}^{(1)}$ that makes an angle $\delta^{(1)}$ with $\vec{k}_{\mathrm{o}}$. However, $v_{\mathrm{ph}}\!=\!c$ along $\vec{k}_{\mathrm{o}}$, but $\widetilde{v}\!=\!\tfrac{c}{\widetilde{n}}\!\neq\!v_{\mathrm{ph}}$ along the same direction ($\widetilde{n}\!<\!1$). (b) A symmetrized superluminal ST wave packet having an X-shaped profile. (c) A subluminal ST wave packet incorporating non-differentiable AD at $\omega\!=\!\omega_{\mathrm{o}}$ (the maximum admissible frequency in the spectrum). Under each panel we plot $\varphi(\omega)$.}
\label{Fig:NonDifferentiableAD}
\end{figure}

\subsection{Tilt angle of the pulse front}

We verified in \cite{Hall21OL} that the pulse-front tilt angle $\delta_{\mathrm{o}}^{(1)}$ for a ST wave packet in the vicinity of $\omega_{\mathrm{o}}$ depends inversely on the square root of the bandwidth $\Delta\omega$ via the ansatz:
\begin{equation}\label{Eq:Ansatz}
\tan{\delta_{\mathrm{o}}^{(1)}}=\frac{\eta}{\sqrt{2\Delta\omega/\omega_{\mathrm{o}}}},
\end{equation}
which provided an excellent fit to the measurements. We are now in a position to offer a derivation of this ansatz and a physical interpretation of the observed effect.

The spectrum of the wave packet cannot extend beyond the non-differentiable frequency $\omega_{\mathrm{o}}$, as illustrated in Fig.~\ref{Fig:NonDifferentiableAD}. Consider a superluminal ST wave packet where $\omega_{\mathrm{o}}$ is the minimum admissible frequency [Fig.~\ref{Fig:NonDifferentiableAD}(a,b)] (a similar analysis applies to the subluminal regime [Fig.~\ref{Fig:NonDifferentiableAD}(c)]), and whose bandwidth $\Delta\omega$ is centered on $\omega_{\mathrm{c}}$ but does \textit{not} include $\omega_{\mathrm{o}}$; i.e., $\omega_{\mathrm{c}}-\omega_{\mathrm{o}}\!>\!\tfrac{1}{2}\Delta\omega$. Making use of Eq.~\ref{Eq:SquareRootOmega}, we have $\omega\varphi^{(1)}\!=\!\tfrac{\eta\omega}{\sqrt{\omega_{\mathrm{o}}(\omega-\omega_{\mathrm{o}})}}$, and substituting $\omega\!=\!\omega_{\mathrm{c}}$ yields:
\begin{equation}\label{Eq:OffAxisADNonDifferentiable}
\tan{\delta_{\mathrm{c}}^{(1)}}=\omega_{\mathrm{c}}\varphi_{\mathrm{c}}^{(1)}=\frac{\eta}{2}\frac{\frac{\omega_{\mathrm{c}}}{\omega_{\mathrm{o}}}}{\sqrt{\frac{\omega_{\mathrm{c}}}{\omega_{\mathrm{o}}}-1}},
\end{equation}
which is bandwidth-independent as usual for TPFs. However, when the spectrum includes the terminus frequency $\omega_{\mathrm{o}}$, the central frequency is $\omega_{\mathrm{c}}\!=\!\omega_{\mathrm{o}}+\tfrac{1}{2}\Delta\omega$, $\tan{\delta_{\mathrm{c}}^{(1)}}\!\approx\!\eta/\sqrt{2\Delta\omega/\omega_{\mathrm{o}}}\!=\!\tan{\delta_{\mathrm{o}}^{(1)}}$ as given in Eq.~\ref{Eq:Ansatz}, and we retrieve the inverse dependence on the square root of the bandwidth, which is a phenomenon unique to ST wave packets.

We can therefore interpret this effect as a consequence of the non-differentiable frequency $\omega_{\mathrm{o}}$ forcibly terminating the spectrum. In the vicinity of $\omega_{\mathrm{o}}$ the AD changes rapidly so that small shifts produce large changes in $\delta_{\mathrm{c}}^{(1)}$ and the pulse-front tilt is sensitive to the bandwidth. When $\omega_{\mathrm{c}}$ is far from $\omega_{\mathrm{o}}$, $\delta_{\mathrm{c}}^{(1)}$ is given by Eq.~\ref{Eq:OffAxisADNonDifferentiable} and the pulse-front tilt no longer depends on the bandwidth.

\subsection{Group-velocity dispersion and higher-order dispersion terms}

Introducing GVD while simultaneously eliminating all higher-order dispersion coefficients requires that $k_{z}^{(2)}(\omega)\!=\!k_{2}$, where $k_{2}$ is a frequency-independent constant and $k_{z}^{(n)}\!=\!0$ for $n\!\geq\!3$. By rewriting Eq.~\ref{Eq:AxialGVDCoefficient} at $\omega$ (rather than at $\omega_{\mathrm{o}}$), $c\omega k_{z}^{(2)}\!=\!\tfrac{d}{d\omega}\{\omega^{2}\tfrac{d\cos{\varphi}}{d\omega}\}$, setting $k_{z}^{(2)}\!=\!k_{2}$, and integrating twice yields $k_{z}(\omega)\!=\!\sigma_{1}+\tfrac{\sigma_{2}}{c}\Omega+\tfrac{1}{2}k_{2}\Omega^{2}$. The integration constants $\sigma_{1}$ and $\sigma_{2}$ can be evaluated using the two boundary conditions at $\omega\!=\!\omega_{\mathrm{o}}$, $k_{z}(\omega_{\mathrm{o}})\!=\!k_{\mathrm{o}}$ and $\tfrac{dk_{z}}{d\omega}\big|_{\omega_{\mathrm{o}}}\!=\!\tfrac{\widetilde{n}}{c}$, which leads to the dispersion relationship $k_{z}(\omega)\!=\!k_{\mathrm{o}}+\tfrac{\widetilde{n}}{c}\Omega+\tfrac{1}{2}k_{2}\Omega^{2}$. We emphasize that this is \textit{not} an approximation in which $k_{z}(\omega)$ is truncated at second-order in $\Omega$; rather, this is an exact dispersion relationship where $\widetilde{n}\!=\!c/\widetilde{v}$, the GVD parameter is $k_{2}$, and \textit{all} terms above second order vanish. From $k_{z}(\omega)$ we extract $\varphi(\omega)$:
\begin{equation}
\sin{\{\varphi(\omega)\}}=\eta\sqrt{\frac{\Omega}{\omega_{\mathrm{o}}}}\;\;\frac{\omega_{\mathrm{o}}}{\omega}\;\;\sqrt{\left\{1+\frac{1+\widetilde{n}}{2}\frac{\Omega}{\omega_{\mathrm{o}}}+\frac{\sigma}{2}\left(\frac{\Omega}{\omega_{\mathrm{o}}}\right)^{2}\right\}\left\{1-\frac{\sigma}{1-\widetilde{n}}\frac{\Omega}{\omega_{\mathrm{o}}}\right\}},
\end{equation}
where $\sigma\!=\!\tfrac{1}{2}k_{2}\omega_{\mathrm{o}}c$. Therefore, realizing a particular value $k_{2}$ for the GVD parameter with all higher-order parameters eliminated requires non-differentiable AD at $\omega_{\mathrm{o}}$ because  $\varphi(\omega)\!\rightarrow\!\eta\sqrt{\tfrac{\Omega}{\omega_{\mathrm{o}}}}$ when $\Omega\!\rightarrow\!0$. A unique feature of non-differentiable AD here is that positive and negative values of $k_{2}$ (normal and anomalous GVD, respectively) are treated on the same footing along the propagation axis in contrast to the result in \cite{Martinez84JOSAA}.

This same approach can be applied to any higher-order axial dispersion coefficient $k_{z}^{(n)}$ with $n\!\geq\!3$. Indeed, it can also be applied to finite superpositions of multiple dispersion terms, thereby making possible the realization of arbitrary dispersion profiles for the first time \cite{Yessenov21ACSP}.

\subsection{Summary of the impact of non-differentiable AD on ST wave packets}

We have found that non-differentiable AD lead to a set of consequences exhibited by ST wave packets that contrast clearly with those listed earlier for differentiable AD in TPFs:
\begin{enumerate}
    \item The group velocity $\widetilde{v}\!=\!c/\widetilde{n}$ can take on arbitrary values along the propagation axis, whereas the phase velocity remains luminal $v_{\mathrm{ph}}\!=\!c$; i.e., $v_{\mathrm{ph}}\!\neq\!\widetilde{v}$ along the propagation axis.
    \item When the spectrum includes the non-differentiable frequency $\omega_{\mathrm{o}}$, the pulse-front tilt angle is bandwidth-dependent. Otherwise, if the spectrum does \textit{not} include the non-differentiable frequency, then the pulse-front tilt angle is bandwidth-independent.
    \item Large deviations of $\widetilde{v}$ from $c$ can be achieved along the propagation axis.
    \item GVD can be altogether eliminated, resulting in a propagation-invariant ST wave packet.
    \item Both normal and anomalous GVD can be realized along the propagation axis and can be treated symmetrically on the same footing.
    \item Higher-order dispersion terms can be isolated and manipulated while suppressing all other dispersion orders along the propagation axis. Indeed, arbitrary dispersion profiles can be readily produced.
\end{enumerate}

\section{Experiment and Results}

\subsection{Setup for non-differentiable AD synthesis}

To demonstrate the impact of non-differentiable AD on the group velocity, we make use of the pulsed-beam shaper we developed recently \cite{Kondakci17NP,Kondakci19NC,Bhaduri19Optica,Yessenov19OPN} for synthesizing ST wave packets; see Fig.~\ref{Fig:Setup}. The arrangement modifies generic plane-wave pulses from a mode-locked Ti:sapphire laser (Tsunami; Spectra Physics) in two steps. In the first step, a diffraction grating (1200~lines/mm) resolves the spectrum of the incident plane-wave pulse and a cylindrical lens (focal length 500~mm) collimates the spectrum. In the second step, a spatial light modulator (SLM; Hamamatsu X10468-02) placed at the focal plane of this lens imparts a two-dimensional phase distribution to the impinging spectrally resolved wave front. This phase distribution is designed so as to deflect each wavelength $\lambda$ independently at a prescribed angle $\varphi(\lambda)$. This facilitates implementing arbitrary AD profiles $\varphi(\lambda)$, whether differentiable \cite{Hall21APLP,Hall21PRA} or non-differentiable \cite{Yessenov21ACSP,Hall21OL,Hall21OL3NormalGVD}. The spatio-temporal spectrum is acquired by a combination of a grating and a lens that produce temporal and spatial Fourier transforms, respectively, and the spatio-temporal wave packet profile is reconstructed by interfering the synthesized wave packet with the original plane-wave pulse (see \cite{Kondakci19NC,Bhaduri19Optica,Yessenov19OE} for details).

\subsection{Non-differentiable AD and the group velocity}

\begin{figure}[t!]
\centering
\includegraphics[width=11cm]{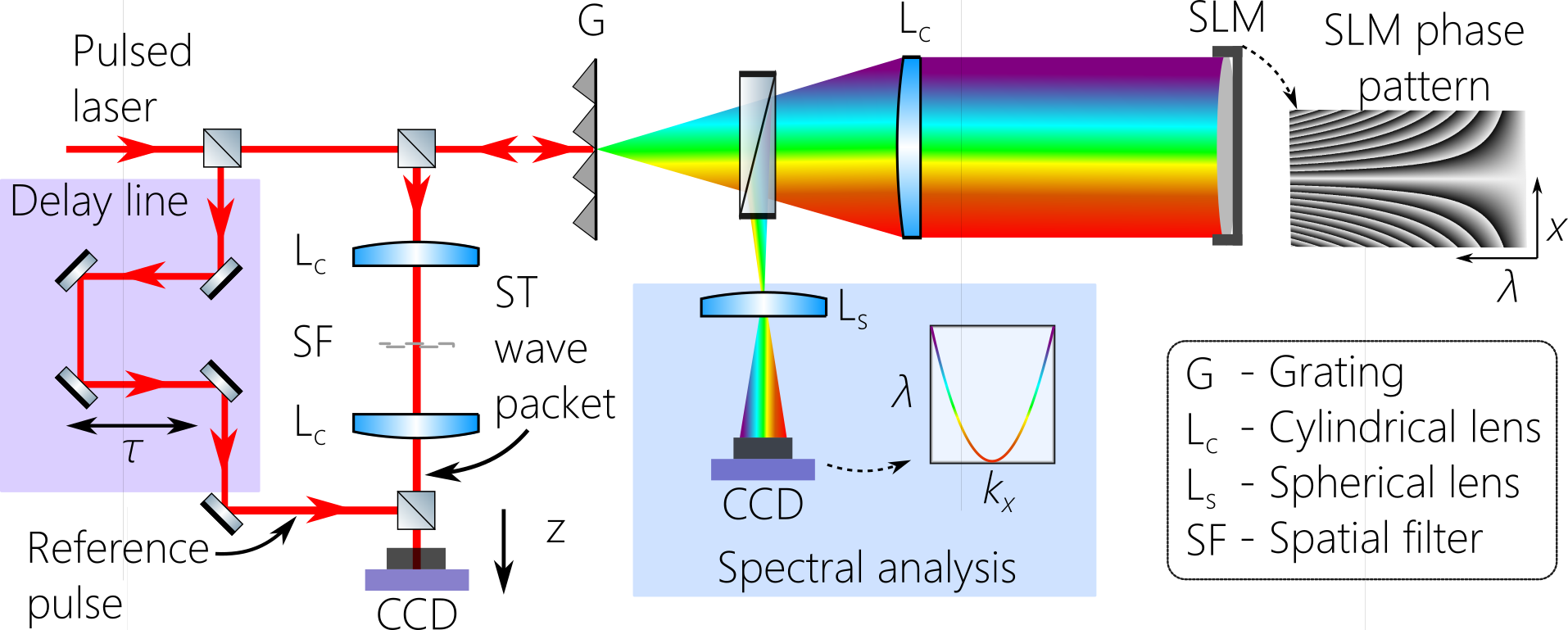}
\caption{Schematic of the experimental arrangement for a pulsed-beam shaper capable of introducing either differentiable or non-differentiable AD into a plane-wave pulse.}
\label{Fig:Setup}
\end{figure}

\begin{figure}[t!]
\centering
\includegraphics[width=13cm]{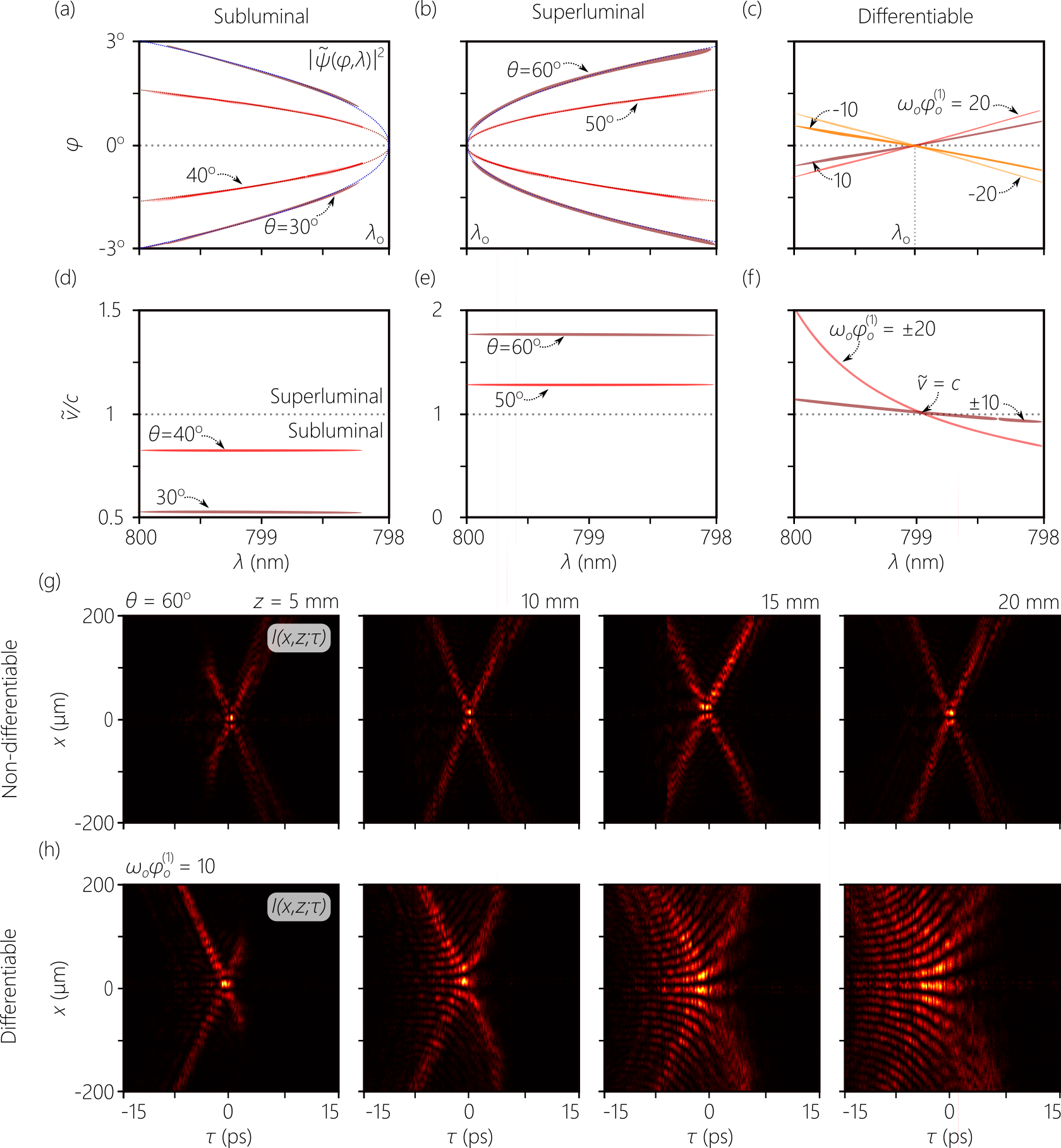}
\caption{(a-c) Measurements of the propagation angle $\varphi(\lambda)$ for (a) subluminal and (b) superluminal ST wave packets endowed with \textit{non-differentiable} AD, and for (c) a TPF endowed with \textit{differentiable} AD. We identify the ST wave packets by their spectral tilt angle $\theta$, where $\widetilde{n}\!=\!\cot{\theta}$ \cite{Kondakci17NP}. (d-f) The group velocity $\widetilde{v}(\lambda)$ extracted from (a-c). (g) The spatio-temporal intensity profile $I(x,z;\tau)$ at different axial planes $z$ for the propagation invariant subluminal ST wave packet from (a), and (h) for the TPF from (e).}
\label{Fig:GroupVelocityData}
\end{figure}

We first demonstrate the impact of the non-differentiability of the AD on the group velocity $\widetilde{v}$. For that purpose, we use the setup in Fig.~\ref{Fig:Setup} to synthesize propagation-invariant subluminal and superluminal ST wave packets endowed with non-differentiable AD and conventional TPFs endowed with differentiable AD. From the measured spatio-temporal spectrum we extract the spectral profile of the propagation angle $\varphi(\lambda)$ and plot the results in Fig.~\ref{Fig:GroupVelocityData}(a) for subluminal wave packets, which fit well the theoretical expectation $\varphi(\omega)\!=\!\eta\sqrt{\tfrac{\Omega}{\omega_{\mathrm{o}}}}$ for small bandwidths. The non-differentiable wavelength $\lambda_{\mathrm{o}}\!=\!798$~nm is the shortest wavelength (highest frequency) admissible in the spectrum. For each value of $\eta\!=\!\sqrt{2(1-\widetilde{n})}$, we extract the wavelength-dependent group velocity $\widetilde{v}(\lambda)$ from $\varphi(\lambda)$ by exploiting the relationship $k_{z}\!=\!\tfrac{\omega}{c}\cos{\varphi}\!=\!k_{\mathrm{o}}+\tfrac{\Omega}{\widetilde{v}}$, from which we have $\widetilde{v}\!=\!c\tfrac{\lambda_{\mathrm{o}}-\lambda}{\lambda_{\mathrm{o}}\cos{\varphi}-\lambda}$. We confirm that $\widetilde{v}\!=\!c/\widetilde{n}\!<\!c$ in each case in Fig.~\ref{Fig:GroupVelocityData}(d). Crucially $\widetilde{v}$ is \textit{independent} of $\lambda$ over the spectrum considered, extending to $\lambda_{\mathrm{o}}$ at $\varphi_{\mathrm{o}}\!=\!0$; that is, the axial group velocity is $\widetilde{v}\!=\!c/\widetilde{n}$ rather than $c$. Similar measurements are carried out in the superluminal regime for propagation-invariant ST wave packets [Fig.~\ref{Fig:GroupVelocityData}(b,e)]. Here the non-differentiable wavelength $\lambda_{\mathrm{o}}\!=\!800$~nm is the longest wavelength (lowest frequency) admissible in the spectrum [Fig.~\ref{Fig:GroupVelocityData}(b)] and $\widetilde{v}\!=\!c/\widetilde{n}\!>\!c$ [Fig.~\ref{Fig:GroupVelocityData}(e)].

For comparison, we plot $\varphi(\lambda)$ for TPFs endowed with differentiable AD in Fig.~\ref{Fig:GroupVelocityData}(c). Using the same experimental configuration but changing the SLM phase distribution to produce $\varphi(\omega)\!\propto\!\Omega$, we find that the associated group velocity $\widetilde{v}(\lambda)$ is \textit{not} independent of $\lambda$ [Fig.~\ref{Fig:GroupVelocityData}(f)]. Crucially, we find that $\widetilde{v}(\lambda_{\mathrm{o}})\!=\!c$, where $\lambda_{\mathrm{o}}\!=\!799$~nm is the wavelength associated with $\varphi_{\mathrm{o}}\!=\!0$; i.e., the on-axis wavelength.

This is further confirmed by acquiring the spatio-temporal profiles along the $z$-axis for a subluminal propagation-invariant ST wave packet incorporating non-differentiable AD corresponding to $\widetilde{n}\!=\!0.58$ [Fig.~\ref{Fig:GroupVelocityData}(g)] to that of a TPF incorporating differentiable AD [Fig.~\ref{Fig:GroupVelocityData}(h)]. The profiles are acquired in moving frames traveling at $c/\widetilde{n}$ and $c$, respectively. In the former the wave packet travels rigidly along $z$, whereas the latter undergoes dispersive temporal broadening.

\begin{figure}[t!]
\centering
\includegraphics[width=13cm]{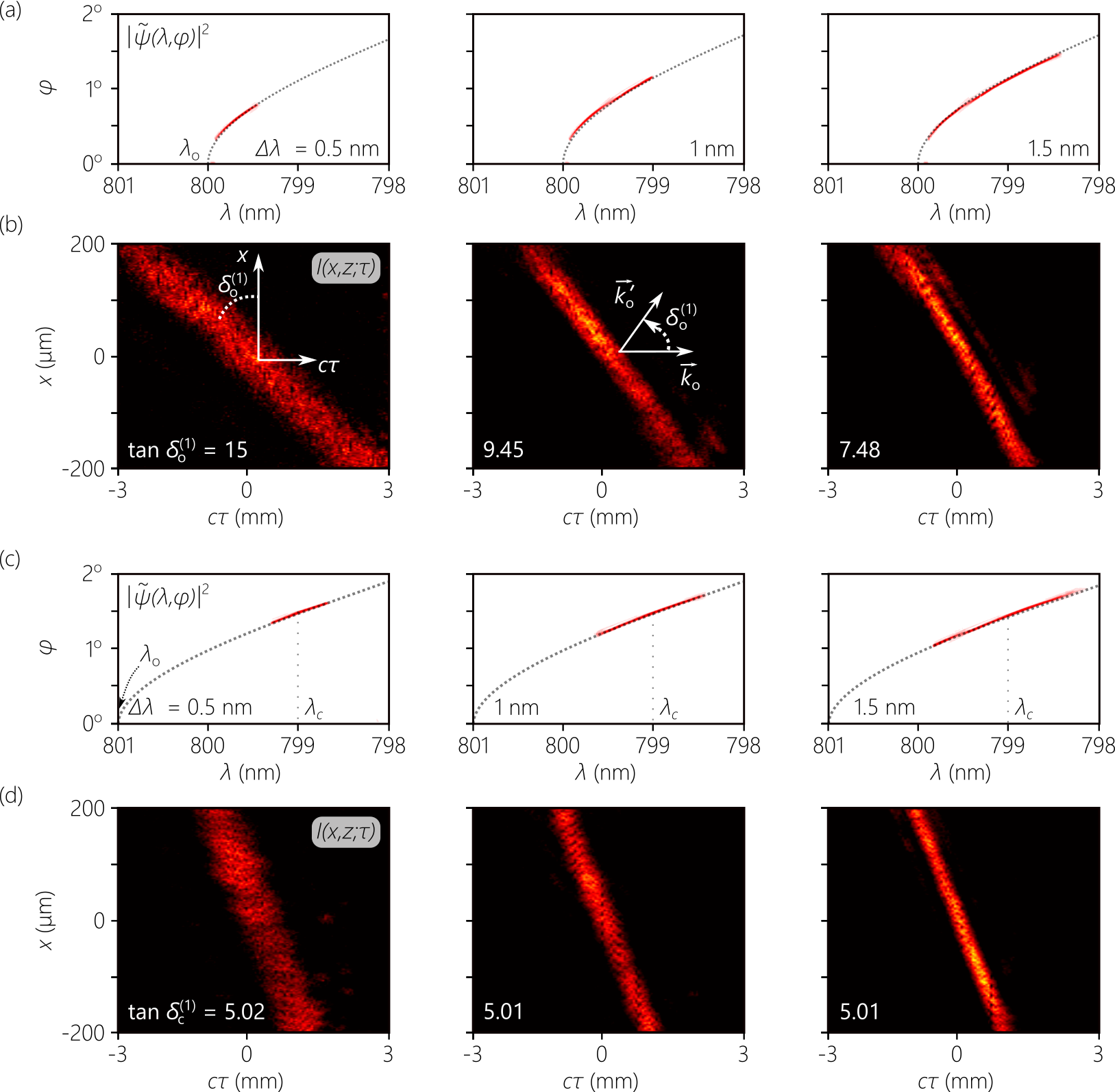}
\caption{Bandwidth-dependence of the pulse-front tilt angle $\delta^{(1)}_{\mathrm{o}}$ in presence of non-differentiable AD. (a) Propagation angle $\varphi(\lambda)$ for a subluminal ST wave packet of different bandwidths $\Delta\lambda$, in the vicinity of the non-differentiable wavelength $\lambda_{\mathrm{o}}\!=\!800$~nm. (b) Spatio-temporal intensity profiles $I(x,0;\tau)$ of the wave packets from (a) demonstrating the change in $\delta^{(1)}_{\mathrm{o}}$ with $\Delta\lambda$. (c,d) Same as (a,b) except that the spectra do not include the non-differentiable wavelength, so that $\delta^{(1)}_{\mathrm{c}}$ evaluated at $\lambda_{\mathrm{c}}\!=\!799$~nm is independent of bandwidth.}
\label{Fig:Bandwidth}
\end{figure}

\subsection{Non-differentiable AD and the pulse-front tilt}

Next, we confirm the impact of non-differentiable AD on the pulse-front tilt angle. We make use of a superluminal ST wave packet ($\widetilde{v}\!\approx\!1.19c$ and $\widetilde{n}\!\approx\!0.84$) whose non-differentiable wavelength is $\lambda_{\mathrm{o}}\!=\!800$~nm. We measure the pulse-front tilt angle as we change the bandwidth from 0.5~nm to 1.5~nm in two scenarios. In the first scenario [Fig.~\ref{Fig:Bandwidth}(a,b)], the bandwidth is measured from $\lambda_{\mathrm{o}}$, so that the central wavelength of the spectrum is $\lambda_{\mathrm{c}}\!=\!\lambda_{\mathrm{o}}-\tfrac{1}{2}\Delta\lambda$, and we observe a clear change in the pulse-front tilt angle with $\Delta\lambda$ in accordance with Eq.~\ref{Eq:Ansatz}. In the second scenario [Fig.~\ref{Fig:Bandwidth}(c,d)], we fix the central wavelength of the wave-packet spectrum at $\lambda_{\mathrm{c}}\!=\!799$~nm and then gradually increase $\Delta\lambda$. Here the spectrum does \textit{not} reach the non-differentiable wavelength $\lambda_{\mathrm{o}}$, and the pulse-front tilt angle is invariant with respect to bandwidth in agreement with Eq.~\ref{Eq:OffAxisADNonDifferentiable}.

\begin{figure}[t!]
\centering
\includegraphics[width=13cm]{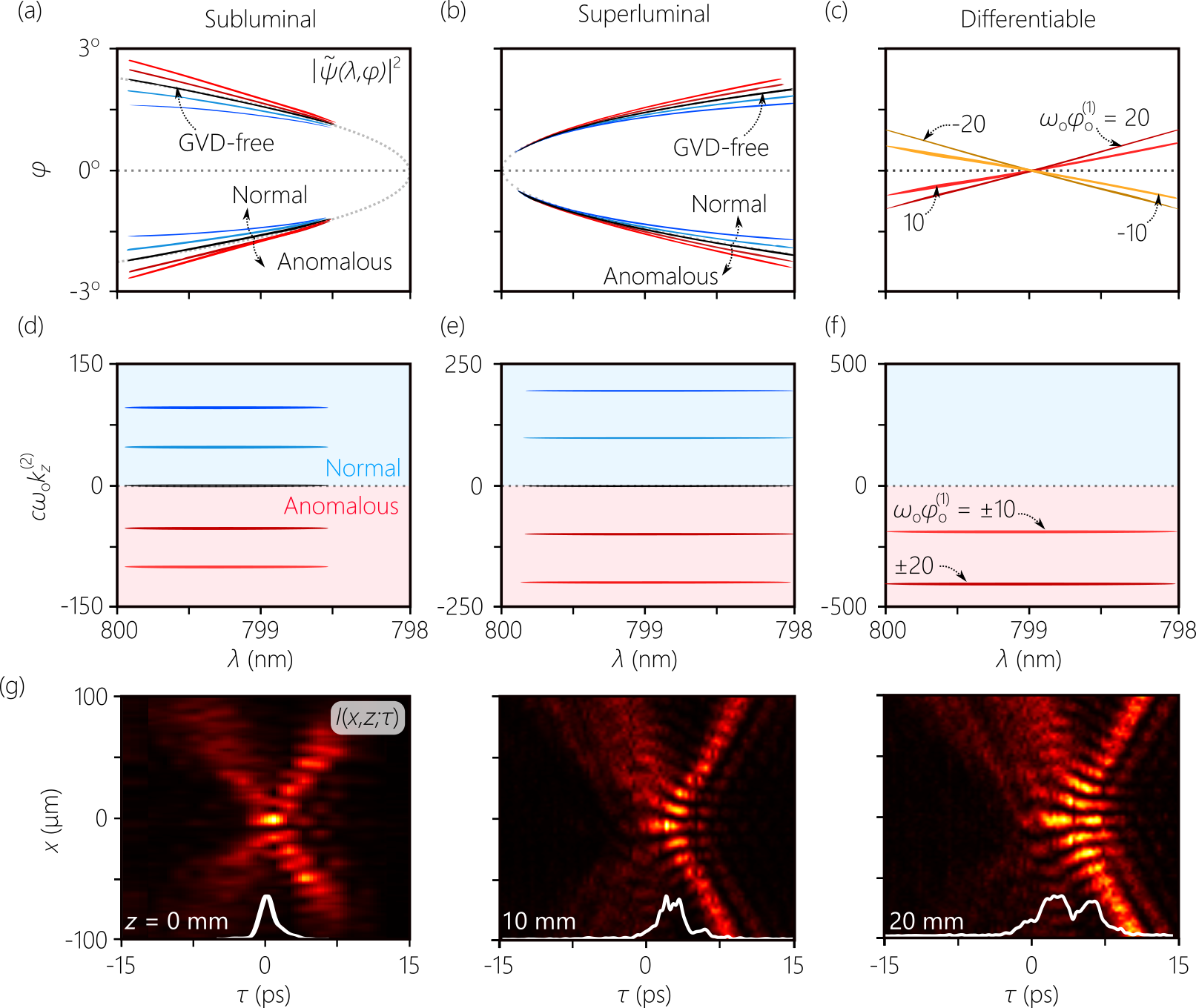}
\caption{Measurements of GVD in presence of non-differentiable AD. (a-c) Measurements of the propagation angle $\varphi(\lambda)$ for (a) subluminal and (b) superluminal ST wave packets endowed with \textit{non-differentiable} AD, and for (c) a TPF endowed with \textit{differentiable} AD. (d-f) The extracted GVD parameter $k_{2}$ from (a-c). (g) The spatio-temporal intensity profile $I(x,z;\tau)$ at different axial planes $z$ for a propagation invariant subluminal wave packet from (a) with normal GVD measured in a frame moving at $c/\widetilde{n}$. The white curve at the bottom of each panel is the pulse profile at the wave packet center $I(0,z;\tau)$.}
\label{Fig:GVDData}
\end{figure}

\subsection{GVD and non-differentiable AD}

Finally, we study the impact of non-differentiable AD on the GVD experienced by ST wave packets in free space in contrast to the GVD exhibited by TPFs endowed with differentiable AD, $\varphi(\omega)\!\propto\!\Omega$. Starting from the propagation-invariant subluminal GVD-free ST wave packet with $\widetilde{n}\!=\!1.19$ from Fig.~\ref{Fig:GroupVelocityData}(a) and a superluminal counterpart with $\widetilde{n}\!=\!0.84$ from Fig.~\ref{Fig:GroupVelocityData}(b), we modify the propagation angle $\varphi(\omega)$ so as to introduce either normal or anomalous GVD. In general, normal GVD requires reducing $\varphi(\omega)$ with respect to the GVD-free counterpart having the same $\widetilde{v}$, and anomalous GVD requires increasing $\varphi(\omega)$.

We plot $\varphi(\omega)$ for both normal and anomalous GVD for the subluminal ST wave packet in Fig.~\ref{Fig:GVDData}(a). We then extract the GVD coefficient and plot it against $\lambda$ in Fig.~\ref{Fig:GVDData}(d). We observe that the extracted GVD parameter $k_{z}^{(2)}\!=\!k_{2}$ is a wavelength-independent constant in both the normal and anomalous regimes, indicating the absence of higher-order dispersion terms. Similar results are plotted in Fig.~\ref{Fig:GVDData}(b,e) for the superluminal ST wave packet. therefore, both normal and anomalous GVD can be produced using either subluminal or superluminal ST wave packets endowed with non-differentiable AD. In the case of differentiable AD with $\varphi(\omega)\!\propto\!\Omega$ [Fig.~\ref{Fig:GVDData}(c)], corresponding to the TPF shown in Fig.~\ref{Fig:GroupVelocityData}(c,f), the resulting GVD is \textit{always anomalous}. The impact of GVD is highlighted in Fig.~\ref{Fig:GVDData}(g) where we plot the spatio-temporal profile along $z$ for a ST wave packet incorporating normal GVD.

\section{Discussion and Conclusion}

We have found that the non-differentiability of $\varphi(\omega)$ at one frequency $\omega_{\mathrm{o}}$ has profound consequences for the free propagation of the optical field. Even if the wave-packet spectrum does \textit{not} include the non-differentiable frequency, it nevertheless exerts influence. An analogous scenario occurs in complex analysis. the function $f(x)\!=\!\tfrac{1}{x^{2}+1}$ is finite and differentiable everywhere on the real line; nevertheless, its Taylor expansion $f(x)\!=\!\sum_{j=0}^{\infty}(-i)^{j}x^{2j}$ converges only in the interval $-1\!<\!x\!<\!1$. The reason is that, when expanded into the complex plane, $f(z)$ does indeed possess singularities along the imaginary axis $z\!=\!\pm i$, which then limits the radius of convergence. In other words, these singularities exert a decisive influence on the behavior of the function $f(x)$ even in domains that do not include these singularities \cite{Needham99book}. Analogously, the non-differentiable frequency $\omega_{\mathrm{o}}$ impacts the behavior of the ST wave packet even when $\omega_{\mathrm{o}}$ does not belong to the spectrum of the wave packet.

Both TPFs and ST wave packets are embodiments of AD. Indeed, these two distinct classes of pulsed beams may have similar spatio-temporal profiles, although their other properties differ from each other in drastic ways. We have identified TPFs as an embodiment of \textit{differentiable} AD inculcated into pulsed fields, whereas ST wave packets are endowed with \textit{non-differentiable} AD. This fundamental distinction undergirds the differences between TPFs and ST wave packets that have been examined here theoretically and experimentally, which we summarize in Table~\ref{Table:Comparison}. It is important to emphasize that the current study has not exhausted all the consequences of introducing non-differentiable AD into a pulsed optical field, and further effects are expected to emerge. Finally, note that X-waves \cite{Lu92IEEEa,Saari97PRL}, which can be considered a particular family of ST wave packets, are unique in that they are AD-free \cite{Yessenov19PRA}.

\begin{table}[t!]
\caption{Summary of the differences between TPFs (endowed with differentiable AD) and ST wave packets (endowed with non-differentiable AD). Here $\vec{k}_{\mathrm{o}}$ is the vector orthogonal to the phase front and represents the propagation direction; $S$ is the wave packet spectrum; $\Delta\omega$: bandwidth (BW); and $\omega_{\mathrm{o}}$ is the non-differentiable frequency.}\label{Table:Comparison}
\begin{tabular}{|p{5cm}||p{4cm}|p{5cm}|}\hline
  & TPFs & ST wave packets \\ \hline\hline
 (1) $v_{\mathrm{ph}}$ and $\widetilde{v}$ along $\vec{k}_{\mathrm{o}}$ & $v_{\mathrm{ph}}\!=\!\widetilde{v}\!=\!c$ & $v_{\mathrm{ph}}\!=\!c$ and $\widetilde{v}\!=\!c/\widetilde{n}$ \\ \hline
 (2) pulse-front tilt angle $\delta_{\mathrm{o}}^{(1)}$& BW-independent & BW-independent when $\omega_{\mathrm{o}}\notin S$; $\tan{\delta_{\mathrm{o}}^{(1)}}\!\propto\!\tfrac{1}{\sqrt{\Delta\omega}}$ when $\omega_{\mathrm{o}}\in S$ \\ \hline
 (3) $\widetilde{v}$ along $\vec{k}_{\mathrm{o}}$ & small deviation from $c$ & large deviation from $c$ \\ \hline
 (4) GVD along $\vec{k}_{\mathrm{o}}$ & always present & can be GVD-free \\ \hline
 (5) sign of GVD along $\vec{k}_{\mathrm{o}}$ & always anomalous & can be normal or anomalous \\ \hline
 (6) high-order dispersion & always present & each order can be tuned independently \\ \hline
\end{tabular}\end{table}

Recent progress in the fabrication of metasurfaces has suggested that the sign and magnitude of the first-order AD term $\omega_{\mathrm{o}}\varphi_{\mathrm{o}}^{(1)}$ can be tuned by varying the structure \cite{Arbabi17Optica,McClung20Light}. It is an open question whether nanostructuring of a single surface can introduce non-differentiable AD into incident pulsed optical fields. A recent study suggests that nonlocal nanophotonic structures may indeed help achieve this task \cite{Guo21Light}.

In conclusion, we have studied the consequences of introducing non-differentiable AD into a pulsed field (ST wave packets) in comparison with the traditional scenario of differentiable AD (TPFs). We have shown that non-differentiable AD provides a unique set of capabilities that are inaccessible to fields produced using conventional optical devices that introduce only differentiable AD. Our theoretical analysis and experiments have revealed that for non-differentiable AD, the following characteristics are exhibited: (1) the on-axis group velocity need not be equal to the phase velocity; (2) the pulse-front tilt angle is bandwidth-dependent when the spectrum on-axis includes the non-differentiable frequency; (3) $\widetilde{v}$ can be readily tuned to values that differ significantly from $c$ over the subluminal, superluminal, and negative-$\widetilde{v}$ regimes; (4) GVD-free ST wave packets can be readily produced; (5) normal and anomalous GVD are treated on the same footing; and (6) higher-order dispersion terms can be isolated and tuned independently. These results dispel some misconceptions that have persisted for decades with regards to what dispersive properties result from AD \cite{Martinez84JOSAA}, and also pave the way to a variety of novel applications in dispersion compensation and nonlinear optics. Furthermore, the utility of these unique characteristics can be extended from freely propagating fields to guided optical modes \cite{Shiri20NC,Guo21PRR} and surface plasmon polaritons \cite{Schepler20ACSP}. 

\section*{Funding}
U.S. Office of Naval Research (ONR) N00014-17-1-2458 and N00014-20-1-2789.

\section*{Disclosures}
The authors declare no conflicts of interest.

\bibliography{diffraction}

\end{document}